\documentstyle[12pt]{article}
\begin{document}
\vspace{1.25in}
\title{Physics with ELFE.
\footnote{Invited talk, International Conference on Nuclear Physics,
Beijing, Aug 21-26, 1995}}
 \vspace{1.in}
\author{{\sl Bernard Frois} \\
{\normalsize DAPNIA, Centre d'Etudes Nucl\'eaires de Saclay,} \\
{\normalsize 91191 Gif sur Yvette, France.}\\
[3mm] {\sl Bernard Pire}\\
[2mm] {\normalsize Centre de Physique Th\'eorique, Ecole Polytechnique,}\\
{\normalsize F-91128 Palaiseau, France}}
\date{\ \ }
\maketitle

\vspace{0.5in} \centerline{{\bf \ Abstract}} \noindent

{ A  15-30 GeV  continuous beam  electron facility  has been proposed by
nuclear physicists in Europe to study how color forces build up  hadrons
from quarks and gluons.  This  project and its physics case are  briefly
reviewed.  The recommendations of NuPECC, the Nuclear Physics  Committee
of the European Science Foundation are presented.}

\section{Introduction}

Recently, NuPECC, the Nuclear Physics Committee of the European  Science
Foundation (P.  Kienle, Chairman) has recommended the construction of  a
15-30 GeV high intensity continuous beam electron accelerator.  The goal
of this  new facility  is to  explore the  quark structure  of matter by
exclusive and semi-inclusive electron scattering from nuclear targets.

In the last  two decades, we  have seen the  emergence of a  theory that
identifies the  basic constituents  of matter  and describes  the strong
interaction  \cite{QCD20}.    The  elementary  building blocks of atomic
nuclei  are  colored  quarks  and  gluons.   The theory describing their
interactions  is  Quantum  Chromodynamics  (QCD)  which  has two special
features, asymptotic freedom and color confinement.  Asymptotic  freedom
means  that  color  interactions  are  weak  at  short distances.  Color
confinement results in the existence of hadrons and in the impossibility
to observe quarks and gluons as single particles.  Color confinement and
asymptotic  freedom  lead  to  the  existence  of two regimes.  At short
distances, quarks and gluons are in the regime of asymptotic freedom and
behave  in  essence  as  free  particles.    At  large  distances, color
interactions are strong and confine quarks and gluons in hadrons (mesons
and  baryons).    A  nucleus  appears  then  to  be  built  of  nucleons
interacting through the exchange of mesons.

Experimental results  show that  {\it at  the fermi  scale} in the dense
nuclear  interior,  nucleons  keep  their  identity.    Thus, a coherent
description of nuclei at that scale has been achieved according to  this
concept of  nuclei made  of nucleons.   At  shorter distances,  nucleons
start  to  overlap  and  one  must  take  their  internal structure into
account.     Mesonic  theory   provides  an   efficient  and  economical
description  of  nuclear  reactions  involving  momentum transfers up to
about 1~(GeV/c)$^2$,  but for  higher momentum  transfers the  situation
becomes much more complex.

The limits of the description of nuclei in terms of nucleons and  mesons
are studied in Europe with the electrons accelerators of Amsterdam, Bonn
and Mainz.  We refer the reader to the reviews given at this  conference
by I. Sick,  T. Walcher and  P. de Witt  Huberts.  In  the United States
nuclear research with  high energy electrons  and photons is  at present
carried out at MIT-Bates accelerator.  In the near future, this research
will be focused at CEBAF, the 4 GeV continuous beam electron accelerator
built  at  Newport  News  (Virginia)  reviewed  at this conference by F.
Gross.   This facility  is nearly  completed.   The accelerator has just
reached its nominal  energy $E =  4$~GeV at the  beginning of 1995.   It
will deliver simultaneously  three beams on  fixed targets.   When CEBAF
was  designed,  the  design  goal  of  superconducting cavities used for
accelerating electrons was  5 MV/m.   Since then, considerable  progress
has been achieved  in the technology  of superconducting cavities.   The
performances of the cavities  delivered by industry exceed  the expected
performances in the  original design of  CEBAF.  Instead  of 4 GeV,  the
final energy of the continuous electron  beam will be of the order  of 6
GeV.  There are already discussions to increase the energy of CEBAF to 8
GeV \cite{CEBAF}.   Beyond 12 GeV  would require essentially  to build a
new facility.

\bigskip

Although one knows the  microscopic theory for the  strong interactions,
{\em one does not understand how quarks build up hadrons}.

\bigskip

After twenty years of  theoretical developments we still  lack reliable,
analytic tools  for this  problem.   This is  one of  the most important
problem of contemporary  physics.  Therefore  a common goal  of nuclear,
particle and astrophysics is today to understand the formation of matter
from quarks and gluons.

High energy data that might shed some light on this problem are  scarce.

\begin{itemize}

\item Deep inelastic  scattering experiments on  nuclei have revealed  a
significant variation  of structure  functions with  the density  of the
nucleus.  This  effect was discovered  by the EMC  collaboration using a
high energy muon beam and subsequently investigated in detail by the NMC
collaboration  also  at  CERN.    Many  different explanations have been
proposed in terms of shadowing, mesons in nuclei, effects of binding  or
modification of the nucleon size in the nuclear medium.

\item Hadron  production at  high transverse  momentum in hadron-nucleus
collisions have revealed a puzzling $A^\alpha $ dependence with  $\alpha
$ varying up to 1.3.  Explanations of this effect involve the successive
scattering  of  a  naked  quark  from  quarks and gluons bound in nearby
nucleons, before the formation of a hadron.

\item Suppression  of charmonium  production has  been observed  in high
energy heavy  ion collisions.   To  isolate the  possible signals from a
quark  gluon  plasma,  it  is  necessary  to  understand the
formation and propagation of a $c\bar c$ pair in a dense medium.

\item Proton-proton elastic scattering  data at large angle  measured at
Brookhaven \cite{CTBNL} seem  to be compatible  with an effect  of color
transparency.  The interpretation  of these data is  still controversial
\cite {RP90}.

\item Diffractive rho-meson production  in muon scattering at  very high
energy  \cite{CTFNAL}  shows  an  increase  in  the production rate from
nuclei at high momentum transfer.  \end{itemize}

\section{\bf The ELFE project}

Electrons  are  pointlike  charges  and  their  interaction  with  other
elementary  particles   is  well   understood.     This  interaction  is
sufficiently weak  to allow  electrons to  penetrate in  the heart  of a
nucleus without  significant perturbation  of its  structure.   Electron
beams  probe  matter  with  a  spatial  resolution that depends on their
energy.  The higher the energy, the better is their resolution.

During  the  last  five  years,  several  conferences and workshops have
discussed the  best experimental  approach to  understand the  evolution
from  quarks  to  hadronic  matter.    Proposals using ELFE (An Electron
Laboratory For Europe):   a $15\div 30$~GeV high  luminosity, continuous
beam electron  accelerator have  been discussed  and collaborations have
been  formed  at  the  Mainz  workshop  in 1992 organized by a committee
composed of J. Arvieux, E. de  Sanctis, T. Walcher (Chairman) and P.  de
Witt Huberts.  This project \cite{ELFE,AP} has been presented to  NuPECC
at the end of 1994.  These proposals form an extensive research  program
on  exclusive  reactions  to  probe  the  evolution of correlated quarks
systems.  Using the nucleus itself  as a microscopic detector is one  of
the important  ideas of  this program.   One  measures the same reaction
using nuclei of different sizes and thus observes the differences in the
evolution from quarks and gluons to hadrons in the nuclear medium.  This
is possible only  in the $15\div  30$~GeV energy range.   One must  have
sufficiently  high  energy  to   describe  the  reaction  in   terms  of
electron-quark scattering.  However,  the energy transfer should  not be
too high since one is interested in the formation of hadrons inside  the
nuclear medium and not outside of the nucleus.

This  research  program  lies  at  the  border  of  nuclear and particle
physics.  Most  of the predictions  of QCD are  only valid at  very high
energies  where  perturbation  theory  can  be  applied.    In  order to
understand  how  hadrons  are  built,  however,  one is in the domain of
confinement where  the coupling  is strong.   Up  to now  there are only
crude theoretical  models of  hadronic structure  inspired by  QCD.  One
hopes that in the next  ten years major developments of  nonperturbative
theoretical methods such as lattice gauge theory will bring a wealth  of
results on the transition  from quark to hadron.   It is fundamental  to
guide   theory   by   the   accurate,   quantitative  and  interpretable
measurements obtained by electron scattering experiments.

The research program of ELFE addresses the questions raised by the quark
structure of matter:   the role  of quark exchange,  color transparency,
flavor  and  spin  dependence  of  structure  functions  and differences
between  quark   distributions  in   the  nucleon   and  nuclei,   color
neutralization in the hadronization of a quark\ldots All these questions
are some of the many exciting facets of the fundamental question:

\begin{center}
{\bf ``How do color forces build up hadrons from quarks and gluons? ''}
\end{center}

ELFE will focus on the following research topics:

\begin{itemize}

\item {\em Exclusive processes}.  Exclusive electroproduction processes,
including  polarization  experiments,  are  needed  to study the spatial
structure of hadrons.  Because  they require coherent scattering of  the
quarks,  exclusive  observables  are  sensitive  to the quark gluon wave
function of the hadrons.  Typical examples are real and virtual  Compton
scattering photo and electroproduction of mesons at large angle and form
factors of mesons or baryons \cite{BP}.

\item {\em Nucleus  as a detector}.   The idea  is  to  use  the  nucleus
as a microscopic
detector  to  determine  the  time  evolution  of  the  elementary quark
configurations in the building up of hadrons.  A typical example of this
research program is color transparency in quasi-elastic reactions and in
charmonium  production.    Another  one  is hadronization in the nuclear
medium.  For these  processes, the nucleus is  used as a medium  of {\em
varying length}.

\item  {\em  Heavy  Flavors}.    The  study  of  the  production and the
propagation of strangeness and charm provides us with an original way to
understand  the  structure  of  hadronic  matter.    The   corresponding
reactions do not involve the valence quarks of the target and probe  its
sea   quark   (intrinsic   strange   or   charm   content)   and   gluon
distributions.

\item  {\em  Short  Range  Structure  of  Nuclei}.   At short distances,
nuclear   structure   cannot   be   reduced   to   nucleons   or  isobar
configurations.    To  unravel  such  exotic  configurations   dedicated
experiments (large  x structure  functions and  $\phi $  production) are
proposed.

\end{itemize}

\section{NuPECC Recommendations}

\begin{enumerate}

\item  NuPECC  has  examined  the  case  for  a  European  CW electron
accelerator  in  the  15-30  GeV  region  (ELFE) which was presented to
NuPECC at the Vienna meeting in April 1994.

\item NuPECC finds the physics  case compelling.  The investigation  of
strongly interacting  systems with  the elementary  probe of  the strong
interaction-the  quark,  produced   in  electron-quark  scattering-   is
essential.   Studies of  hadron structure  by exclusive  experiments are
indispensable  for  a  better  understanding  of QCD in the confinement
regime.  New windows for the investigation of hadronic matter are opened
by the use of probes with strangeness and charm.

\item NuPECC considers the potential application of highest  brilliance
ultrarelativistic beams in the  production of coherent short  wavelength
radiation of high intensity to be very promising.

\item NuPECC recommends  that appropriate action  is taken in  order to
proceed  towards  the  construction  of  a  European  facility providing
electron beams of high duty cycle and brilliance in the 15-30 GeV energy
range.    It  should  serve  scientists  from  universities and research
laboratories as a central users facility.

\end{enumerate}

In order to reach this goal, the following steps need to be taken.

\begin{description}

\item[a)]  Substantial  advancement   of  the  state   of  the  art   of
superconducting RF and cryotechnology is necessary to construct such  an
accelerator in a cost effective way.  NuPECC recognizes that significant
progress has already  been made.   It sees important  synergies with the
development needed for linear colliders under consideration by  particle
physicists.  During the next few  years, an intense joint effort by  the
European laboratories  involved is  required to  develop the  technology
further.    The  technical   feasibility  of  the  crucial   accelerator
components should be established at a testbed facility.

\item[b)] An ELFE coordinating group should be formed by the  scientific
community  from  the  field  of  interest,  with  the  help of NuPECC if
required.    This  group  should  coordinate  the technical developments
needed,   and   integrate   the   present   experimental   programs   in
electromagnetic  physics  in  order  to  create an enlarged and coherent
community  in  preparation  for  the  long  range  future with ELFE.  An
important aspect of  this the R\&D  work for the  experimental equipment
required for the physics proposed.

\item[c)] It is important to develop further potential applications.  In
particular  those   based  on   coherent  radiation   produced  by   the
high-brilliance beams.

\end{description}

A few weeks before this conference,  a meeting has been organized by  S.
Bass   in   Cambridge   (U.K.)   to   discuss  how  to  implement  these
recommendations in  the present  European context.   The  possibility of
using a  superconducting linear  accelerator injecting  electrons in the
HERA ring used  as a stretcher  has been discussed  by Brinkmann.   This
fall  a  group  of   European  accelerator  physicists  will   start  to
investigate this possibility.  At its next meeting in September,  NuPECC
is expected to propose the formation of an initiative group to  organize
the work between European physicists interested in this project.

\section{\bf Exclusive reactions:  A new tool}

Exclusive reactions are processes in which the final state is completely
resolved.  They  are important since  at high momentum  transfers, they
are sensitive to the quark composition themselves as expressed by  quark
distribution amplitudes.

To probe hadronic structure at  very small distances $\lambda $  we must
transfer to the target a  large momentum transfer $Q=1/\lambda  $.
The recoil kinetic energy being usually larger then the rest mass of the
hadron, one is  led to use  a relativistic framework.   Feynman was  the
first to show that  by looking at the  hadron from an infinite  momentum
frame,  one   develops  an   intuitive  understanding   of  relativistic
collisions.  This  leads to view  hadrons as a  collection of quasi-free
objects, the partons, sharing each a fraction $0<x_i<1$ of the  infinite
momentum and  moving closely  parallel to  it.   They are  bound by  the
strong color  force but  their binding  energy being  small compared  to
their momentum,  they behave  almost freely.   The  parton model  is for
photon-hadron interactions what the impulse approximation is for nuclei.
The major difference is that partons cannot be directly observed due  to
the existence of confinement.

Nearly all  existing data  on quark  distributions in  hadrons have been
obtained by {\em \ inclusive  scattering} of high energy particles.   In
such reactions, one strikes quarks with considerable momentum and energy
and  reconstructs  quark  distributions  from  scattering data.  This is
possible  because  of  a  property  of  factorization  of the scattering
amplitudes in quantum field theory.  This property has allowed theory to
find a  firm basis  for the  partonic description  and to  go beyond the
original  model  proposed  by  Feynman  and  Bjorken.   The experimental
observation  amounts  to  an   average  over  all  the   possible  quark
configurations in the nucleus.  In addition to fundamental tests of QCD,
the measurements of  structure functions have  lead to the  discovery of
the importance of gluons in  the momentum and spin distributions  in the
proton.We  now  need  to  go  further  and  understand  how simple quark
configurations  are  controlled  by  confining  mechanisms.  One needs a
different type of data  sensitive to the time  evolution of a system  of
correlated quarks.   This  is the  domain of  exclusive reactions  where
scattered  particles  emitted  in  a  specific  channel  are observed in
coincidence.

In {\em  exclusive reactions  }\cite{EXCLTHEORY}, one  first writes  the
wave function of a composite state as a simple expansion of Fock  states
with a fixed number of quarks and gluons :

\begin{equation}
\left| \pi \right\rangle =\Psi _v\left| q\overline{q}\right\rangle +\Psi
_g\left| q\overline{q},g\right\rangle +\Psi _{q\overline{q}}\left| q
\overline{q},q\overline{q}\right\rangle
\end{equation}

\begin{equation}
\left| N\right\rangle =\Psi _V\left| qqq\right\rangle +\Psi _g\left|
qqq,g\right\rangle +\Psi _{q\overline{q}}\left| qqq,q\overline{q}
\right\rangle +...
\end{equation}

\bigskip
\noindent
where the valence component $\Psi _V$ turns out to be the dominant one in
exclusive reactions.

Here the quarks are ``current'' quarks and not ``constituent'' ones.   A
constituent quark  may be  seen as  a complex  structure consisting of a
current  quark   ``dressed''  of   quark-antiquark  pairs   and  gluons.
Constituent quarks  are important  to get  an intuitive  picture of  the
quark structure of  the nucleon, but  they cannot be  used to understand
quark dynamics in the framework of a relativistic quantum field theory.

The  valence  wave  functions  $\Psi  _V$  are  functions  of light-cone
momentum fraction $x_i$, transverse momentum $p_T$ and helicities.  They
contain  important  information  on  quark  confinement  dynamics.    By
integrating $\Psi _V$ over transverse momenta, one gets the distribution
amplitude $\phi (x_i) $.

The analysis of leading QCD  corrections to any exclusive amplitude  has
shown  \cite{EXCLTHEORY}  that  these  distribution  amplitudes  obey  a
renormalization group equation, leading  to a well understood  evolution
in terms  of perturbative  QCD.   At asymptotic  $Q^2$, the distribution
amplitudes simplify, e.g. for the proton:
\begin{equation}
\phi (x_1,x_2,x_3,Q^2)\rightarrow x_1x_2x_3\delta (1-x_1-x_2-x_3)
\end{equation}

The $Q^2$ evolution  is however sufficiently  slow for the  distribution
amplitude  to  retain  much   information  at  measurable  energies   on
confinement physics.  The experimental strategy of ELFE physics is  thus
to sort out  the hadron distribution  amplitudes from various  exclusive
reactions to learn about the dynamics of confinement.  This is  possible
thanks  to  the  fact  that,  within  perturbative  QCD,  one  derives a
factorization property of exclusive  scattering amplitudes which may  be
schematically written as :

\begin{equation}
{\cal M}=\phi (x_i,Q^2)\otimes T_H(x_i,y_i,Q^2)\otimes \phi ^{*}(y_j,Q^2)
\end{equation}
where integrals  over momentum fractions  $x_i$ and $y_j$  are
implicit.  The hard scattering $T_H$ is calculable perturbatively as  an
expansion in $\alpha _s(Q^2)$ free of large logarithmic corrections.
The functions $\phi _i$ are the non perturbative distribution amplitudes
describing the valence quark content of the proton.

One  may  ask  the  question  whether  existing  high  energy   electron
accelerators designed to study electroweak physics give access to  these
distribution amplitudes.  This is not possible because of the  smallness
of exclusive amplitudes at  large transfers. Let us illustrate this point by a
back of the envelope order of magnitude
 estimate ; take $Z_0$ decays as measured in great details at LEP.
 From the known decay rate into an electron-positron pair (around 3 per cent)
>and
the counting rules for meson form factors\cite{EXCLTHEORY}, one infers that the
decay rate to an exclusive light meson pair is less than one billionth.
 Accelerators  designed for studying electroweak physics or QCD  in inclusive
 reactions do not  give us an access to the dynamics of confinement.

The only  possibility is  to use  a dedicated  high intensity continuous
beam  electron  accelerator  to  study  exclusive  reactions  at   large
transfer.

\section{The Nucleus as a femto-detector}

A  central  idea  of  the  ELFE  project  is  to  use  the  nucleus as a
microscopic detector to determine  the time evolution of  the elementary
quark  configurations  in  the  building  up  of  hadrons.   Two typical
examples   of   this   research   program   are  color  transparency  in
quasi-elastic  (e,e'p)  reactions  and  in  charmonium  production,  and
hadronization in the nuclear medium.   For these processes, the  nucleus
is used as a medium of {\em varying length}.

The typical time scales to build up a hadron is $\tau _o\sim 1$ fm/c  in
its rest frame.   This is the time  needed by a quark  to travel through
distances  characteristics  of  confined  systems.    Due to the Lorentz
dilation  factor  $  \gamma  =E/M$,  the  time  scale  $\tau  $,  in the
laboratory frame, is several fm/c's.

\begin{center}
{\em At this scale, the only available detector is the nucleus.}
\end{center}

Color  transparency  has  been  recently extensively discussed\cite{CT}.
This phenomenon illustrates the power of exclusive reactions to  isolate
simple elementary quark configurations.   The experimental technique  to
probe these configurations is the following:

\begin{itemize}

\item For  a hard  exclusive reaction,  say electron  scattering from  a
proton, the  scattering amplitude  at large  momentum transfer  $Q^2$ is
suppressed  by  powers  of  $Q^2$  if  the proton contains more than the
minimal number  of constituents.   This  is derived  from the  QCD based
quark  counting   rules,  which   result  from   the  factorization   of
wave-function-like  distribution  amplitudes.    Thus protons containing
only  valence  quarks  participate  in  the  scattering.  Moreover, each
quark,  connected  to  another  one  by  a  hard gluon exchange carrying
momentum of  order $Q$,  should be  found within  a distance  of order $
1/Q$.    Thus  ,  at  large  $Q^2$  one  selects  a  very  special quark
configuration:  all connected quarks are close together, forming a small
size color neutral  configuration sometimes referred  to as a  {\em mini
hadron}.   This mini  hadron is  not a  stationary state  and evolves to
build up a normal hadron.

\item Such  a color  singlet system  cannot emit  or absorb  soft gluons
which carry energy or momentum smaller than $Q$.  This is because  gluon
radiation --- like photon radiation in QED --- is a coherent process and
there is thus destructive interference between gluon emission amplitudes
by quarks  with ``opposite''  color.   Even without  knowing exactly how
exchanges  of   soft  gluons   and  other   constituents  create  strong
interactions, we  know that  these interactions  must be  turned off for
small color singlet objects.

\end{itemize}

An exclusive hard reaction will thus probe the structure of a {\em  mini
hadron}, i.e. the short distance part of a minimal Fock state  component
in the hadron  wave function.   This is of  primordial interest for  the
understanding of the difficult physics of confinement.  First, selecting
the simplest Fock state amounts to the study of the confining forces  in
a   colorless   object   in   the   ''quenched   approximation''   where
quark-antiquark pair creation from  the vacuum is forbidden.   Secondly,
letting the mini-state evolve during its travel through different nuclei
of various  sizes allows  an indirect  but unique  way to  test how  the
squeezed mini-state  goes back  to its  full size  and complexity,  {\em
i.e.} how  quarks inside  the proton  rearrange themselves  spatially to
''reconstruct'' a normal size hadron.   In this respect the  observation
of baryonic resonance  production as well  as detailed spin  studies are
mandatory.

To the extent that the electromagnetic form factors are understood as  a
function of $Q^2$, $eA  \rightarrow e^{\prime}(A-1) p$ experiments  will
measure  the  color  screening  properties  of  QCD.  The quantity to be
measured is the transparency ratio $T_r$ which is defined as:

\begin{equation}
T_r = \frac{\sigma_{Nucleus}}{Z \sigma_{Nucleon}}
\end{equation}

At asymptotically large values  of $Q^2$, dimensional estimates  suggest
that $ T_r$ scales as a function of $A^{\frac 13}/Q^2$.  The approach to
the scaling behavior as well as the value of $T_r$ as a function of  the
scaling   variable   determine   the   evolution   from   the  pointlike
configuration to the  complete hadron.   This highly interesting  effect
can   be   measured   in   quasieleastic   electron   proton  scattering
$(e,e^{\prime  }p)$  reaction  that  provides  the  best  chance  for  a
quantitative interpretation.

A  first  experiment  at  SLAC  (NE-18)  \cite{CTSLAC}  has  performed a
preliminary exploration of Color Transparency with the (e,e'p)  reaction
for H,  C, Fe  and Au,  in the  $Q^2$ range  of 1--7~GeV$^2$.   In these
kinematics, no effect was observed.

\section{Accelerator and detectors}

The choice of the energy range of 15 to 30 GeV for the ELFE  accelerator
is fixed by three constraints:

\begin{itemize}

\item Hard electron-quark scattering:   one must have  sufficiently high
energy  and  momentum  transfer  to  describe  the  reaction in terms of
electron-quark scattering.  The high  energy corresponds to a very  fast
process where the struck quark  is quasi-free.  High momentum  transfers
are necessary to probe short distances.

\item  Nuclear  sizes:    The  energy  of  the incident electron beam is
determined to match the characteristic  interaction time $\tau $ to  the
diameter of the nucleus.  Starting from the rest frame time $\tau _o\sim
1$ fm/$c$  and taking  into account  a typical  Lorentz dilation  factor
$\gamma =E/M$  this means  a time  $\tau $  of several  fm/$c$'s in  the
laboratory.   If the  energy transfer  is too  large, the building-up of
hadrons occurs outside the nucleus which can then no longer be used as a
microscopic detector.

\item Charm production requires a minimum electron beam energy of 15 GeV
to have reasonable counting rates.  \end{itemize}

\begin{table}[ht]
\begin{center}
\begin{tabular}{|l|r|}
\hline\hline
Beam Energy & $15 \div 30 GeV$ \\
Energy Resolution FWHM & $3 \times 10^{-4}$ @ 15 GeV \\
& $10^{-3}$ @ 30 GeV \\
Duty Factor & $\simeq 100$~\% \\
Beam Current & $10 \div 50 \mu$A \\
Polarized Beams & $P > 80$~\% \\ \hline\hline
\end{tabular}
\end{center}
\caption{ELFE Accelerator Parameters}
\label{t31}
\end{table}

Exclusive and semi-inclusive  experiments are at  the heart of  the ELFE
project.  To avoid a prohibitively large number of accidental coincident
events a high duty cycle  is imperative.  The ELFE  experimental program
also  requires  a  high   luminosity  because  of  the   relatively  low
probability of exclusive processes.  Finally a good energy resolution is
necessary to identify specific reaction channels.  A typical  experiment
at 15  GeV (quasielastic  scattering for  instance) needs  a beam energy
resolution of about  5 MeV. At  30 GeV the  proposed experiments require
only  to  separate  pion  emission.    These characteristics of the ELFE
accelerator are summarized in table \ref{t31}.

Due  to  the  very  low  duty  cycle  available at SLAC and HERA (HERMES
program)  one  can  only  perform  with  these  accelerators   inclusive
experiments and a limited set of exclusive experiments.

\begin{center}

{\em ELFE  will be  the first  high energy  electron beam  beyond 10 GeV
\\with both high intensity and high duty factor.}

\end{center}

The  various  components  of  the  ELFE experimental physics program put
different requirements on  the detection systems  that can be  satisfied
only  by  a  set  of  complementary  experimental  equipment.   The most
relevant detector features are  the acceptable luminosity, the  particle
multiplicity, the angular acceptance and the momentum resolution.   High
momentum   resolution   ($5   \times   10^{-4}$)   and  high  luminosity
($10^{38}$~nucleons/cm$^2$/s)  can  be  achieved  by  magnetic  focusing
spectrometers.   For semi-exclusive  or exclusive  experiments with more
than two  particles in  the final  state, the  largest possible  angular
acceptance  ($\sim  4  \pi$)  is  highly  desirable.    The  quality and
reliability of large acceptance detectors have improved substantially in
the last two decades.  The design of the ELFE large acceptance detectors
uses state of  the art developments  to achieve good  resolution and the
highest possible luminosity.

\section{CONCLUSIONS}

The ELFE  research program  lies at  the border  of nuclear and particle
physics.  Most  of the predictions  of QCD are  only valid at  very high
energies  where  perturbation  theory  can  be  applied.    In  order to
understand  how  hadrons  are  built,  however,  one is in the domain of
confinement where the  coupling is strong.   It is  fundamental to guide
theory  by  the  accurate,  quantitative  and interpretable measurements
obtained by electron scattering experiments, in particular in  exclusive
reactions.

{\em This research domain is  essentially a virgin territory.   There is
only a limited amount of experimental data with poor statistics.  It  is
not possible to  make significant progress  in the understanding  of the
evolution from quarks to hadrons with the available information.}

This  lack  of  data  explains  to  a  large  extent  the  slow  pace of
theoretical progress.   The situation will  considerably improve due  to
technical breakthroughs in electron accelerating techniques.  ELFE  will
be the first high energy  machine offering the high luminosity  and high
duty cycle demanded by the exclusive reaction program.

A few topics of the experimental program proposed at ELFE can be covered
by existing or planned facilities at the price of considerable  efforts.
This is the case of the proton  electric form factor at SLAC.  Also  the
proton transverse spin structure function can be studied at RHIC through
dilepton pair production in  polarized proton-proton collisions.   These
topics are but a small part of the extensive ELFE research program.  The
exploratory program on color transparency at Brookhaven with protons and
at SLAC with electrons did strengthen the need for dedicated experiments
with high  energy resolution  and high  duty cycle  electron beam.   The
HERMES program at HERA proposes a first detailed study of semi-inclusive
reactions.  ELFE experiments will  increase the statistics by orders  of
magnitude  thus  allowing  a  much  more detailed understanding of color
neutralization.

\bigskip

The goal of the ELFE research program, starting from the QCD  framework,
is to explore the coherent and quark confining QCD mechanisms underlying
the strong force.  It is not to test QCD in its perturbative regime, but
rather to use  the existing knowledge  of perturbative QCD  to determine
the reaction mechanism and access the hadron structure.

\vspace*{0.5cm} \noindent

{\em ELFE will use the tools that have been forged by twenty years of
research in QCD, to elucidate the central problem of color interaction:
color confinement and the quark and gluon structure of matter.}

\bigskip

{\Large {\bf Ackowledgements}}

\vspace*{0.5cm}

We are  most grateful  to our  colleagues from  China who have organized
this conference for their kind  invitation to present the ELFE  project.
This overview is based on  the original research proposals for  the ELFE
project\cite{ELFE,AP}.  The ``Centre de Physique Th\'eorique'' of  Ecole
Polytechnique, is  a division  of CNRS,  the French  Center for National
Scientific Research.


\begin{thebibliography}{99}

\bibitem{QCD20} QCD  20 years  later, edited  by P.M.   Zerwas  and H.A.
Kastrup, (World Scientific, Singapore, 1993)

\bibitem{CEBAF} Workshop on CEBAF at Higher Energies, edited by N. Isgur
and P. Stoler, CEBAF 1994.


\bibitem{CTBNL}  A.S. Carroll {\it et\ al} Phys. Rev. Lett. {\bf 61}, 1698
(1988).


\bibitem{RP90} J.P. Ralston and B. Pire, Phys.  Rev.  Lett.  {\bf  65},
2343 (1990).

\bibitem{CTFNAL} M.R.Adams {\it et\ al}, Phys.  Lett.B (to be published)

\bibitem{ELFE} The ELFE PROJECT, edited by  J. Arvieux and E. De  Sanctis
(Italian Physical Society, Bologna, 1993)

\bibitem{AP} J.  Arvieux and  B. Pire,  Progress in
Particle and Nuclear Physics {\bf 30},
 299 (1995).

\bibitem{BP} T. Gousset and B. Pire,  proceedings of the ELFE summer school on
confinement physics, Cambridge (1995), edited by S. Bass (Editions Frontieres,
 Gif-sur-Yvette, France 1996).

\bibitem{EXCLTHEORY} S.J.  Brodsky and G. R. Farrar, Phys Rev Lett  {\bf
31} (1973) 1153.  V. A.  Matveev, R. M. Muradyan and A.  V. Tavkhelidze,
Lett.   Nuovo Cimento  {\bf 7},  719 (1973).   S.  J. Brodsky  and G. P.
Lepage, Phys.   Lett.  {\bf  87}B, 359 (1979),  A. V. Efremov  and A. V.
Radyushkin, Phys.  Lett.   {\bf 94}B, 245(1980),  V. L. Chernyak, V.  G.
Serbo and A. R. Zhitnitsky, Yad.  Fiz.  {\bf 31}, 1069 (1980), A. Duncan
and A. H. Mueller, Phys.  Rev.  D {\bf 21 }, 1636 (1980), S. J.  Brodsky
and G. P. Lepage, Phys.  Rev.   D {\bf 22}, 2157 (1980).  For a  review,
see S. J. Brodsky and G. P. Lepage in {\it Perturbative QCD}, edited  by
A. H. Mueller (World Scientific, Singapore, 1989).

\bibitem{CT} A. Mueller, {\it Proceedings of the Seventeenth Recontre de
Moriond on Elementary Particle Physics}, (Les Arcs, France 1982)  edited
by J. Tran Thanh Van (Editions Frontieres, Gif-sur-Yvette, France 1982),
p.  13;  S.   J.  Brodsky,  in   {\it  Proceedings  of   the  Thirteenth
International Symposium on Multiparticle Dynamics}, edited by W. Kittel,
W. Metzger, and A. Stergiou (World Scientific, Singapore, 1982) p.  963.
see also L. Frankfurt, G.A.  Miller, M. Strikman, Comments Nucl.   Part.
Phys.    {\bf  21},  1  (1992), P.Jain, B. Pire and J.P. Ralston, to be
published in Phys. Rep. (1996)  and  the  numerous  contributions to the
proceedings of  PANIC XIII,  edited by  A. Pascolini  (World Scientific,
Singapore, 1994).

\bibitem{CTSLAC}  T. A. Armstrong {\it et al.}, Phys. Rev. Lett. {\bf 70},
1212 (1993).
\end{thebibliography}
\end{document}